\documentclass[aps,twocolumn,prl,showpacs,amsmath,amssymb,10pt]{revtex4-2}
\usepackage{graphicx}
\usepackage{scrextend}
\usepackage{manfnt,pifont}
\usepackage{hyperref}
\usepackage{url}
\usepackage{lipsum}
\usepackage[font=small]{subcaption}
\usepackage[font=small]{caption}

\usepackage[normalem]{ulem}

\usepackage{url}
\usepackage[outline]{contour}
\usepackage{textgreek}

\usepackage{tikz}
\usetikzlibrary{shapes}

\usepackage{CJK}
\usepackage{enumitem}
\setlist[itemize]{leftmargin=*}

\newcommand\wh\widehat

\makeatletter
\DeclareFontFamily{OMX}{MnSymbolE}{}
\DeclareSymbolFont{MnLargeSymbols}{OMX}{MnSymbolE}{m}{n}
\SetSymbolFont{MnLargeSymbols}{bold}{OMX}{MnSymbolE}{b}{n}
\DeclareFontShape{OMX}{MnSymbolE}{m}{n}{
    <-6>  MnSymbolE5
   <6-7>  MnSymbolE6
   <7-8>  MnSymbolE7
   <8-9>  MnSymbolE8
   <9-10> MnSymbolE9
  <10-12> MnSymbolE10
  <12->   MnSymbolE12
}{}
\DeclareFontShape{OMX}{MnSymbolE}{b}{n}{
    <-6>  MnSymbolE-Bold5
   <6-7>  MnSymbolE-Bold6
   <7-8>  MnSymbolE-Bold7
   <8-9>  MnSymbolE-Bold8
   <9-10> MnSymbolE-Bold9
  <10-12> MnSymbolE-Bold10
  <12->   MnSymbolE-Bold12
}{}

\let\llangle\@undefined
\let\rrangle\@undefined
\DeclareMathDelimiter{\llangle}{\mathopen}%
                     {MnLargeSymbols}{'164}{MnLargeSymbols}{'164}
\DeclareMathDelimiter{\rrangle}{\mathclose}%
                     {MnLargeSymbols}{'171}{MnLargeSymbols}{'171}
\makeatother

\usepackage{eso-pic}
\newcommand{\reportnum}[2]{
  \AddToShipoutPictureBG*{%
    \AtPageUpperLeft{%
      \hspace{0.75\paperwidth}%
      \raisebox{#1\baselineskip}{%
        \makebox[0pt][l]{\textnormal{#2}}
  }}}%
}

\begin{document}

\begin{CJK*}{UTF8}{GB}
\CJKfamily{gbsn}

\reportnum{-3}{USTC-ICTS/PCFT-24-25}

\title{Correlators of $\mathcal{N}=4$ SYM on real projective space at strong coupling}

\author{Xinan Zhou$^{a,b}$}
\affiliation{$^{a}$Kavli Institute for Theoretical Sciences, University of Chinese Academy of Sciences, Beijing 100190, China}
\affiliation{$^{b}$Peng Huanwu Center for Fundamental Theory, Hefei, Anhui 230026, China}

\begin{abstract}
We consider type IIB supergravity on a $\mathbb{Z}_2$ quotient of AdS$_5\times$S$^5$ as the holographic dual of strongly coupled 4d $\mathcal{N}=4$ SYM on $\mathbb{RP}^4$ space with the gauging of charge conjugation. Using bootstrap techniques, we determine all two-point functions of $\frac{1}{2}$-BPS operators of arbitrary weights at the leading order in the large central charge expansion.
\end{abstract}

	\maketitle
\end{CJK*}

\noindent {\bf Introduction.} Placing theories on nontrivial backgrounds is interesting because it gives access to features which are otherwise invisible in infinite flat space. A simple example is CFTs with a planar conformal boundary. The boundary partly breaks conformal symmetry but also introduces infinitely many new data. These are associated with the new operators living on the boundary, and consist of their dimensions, OPE coefficients, as well as two-point function coefficients with operators in the bulk. In this paper, we are interested in putting CFTs on real projective space. Real projective space is defined by the identification under conformal inversion $x^\mu\leftrightarrow -x^\mu/x^2$ and is the simplest non-orientable manifold in even dimensions. Defining such theories is possible for any QFT with time-reversal symmetry and is of interest in both condensed matter and high energy physics to study subtle anomalies involving time reversal \cite{Hsieh:2015xaa,Witten:2015aba,Metlitski:2015yqa,Seiberg:2016rsg,Witten:2016cio,Tachikawa:2016cha,Tachikawa:2016nmo,Barkeshli:2016mew,Guo:2017xex,Wan:2019oyr,Wang:2019obe} . Moreover, for CFTs real projective space also turns out to be a useful arena for the conformal bootstrap  \cite{Nakayama:2016cim,Hasegawa:2016piv,Hogervorst:2017kbj,Hasegawa:2018yqg,Giombi:2020xah}. 

We are particularly interested in $\mathcal{N}=4$ SYM with $SU(N)$ gauge group placed on $\mathbb{RP}^4$ in a way preserving half of supersymmetry \cite{Wang:2020jgh,Caetano:2022mus}. This is motivated by the following reasons. The paradigmatic $\mathcal{N}=4$ SYM theory is the arguably the simplest yet nontrivial theory
that is integrable, amenable to supersymmetric localization, and is the prime example of AdS/CFT. It is very interesting to see how these techniques extend to the case of real projective space, especially in the strongly coupled regime. On the other hand, among the nontrivial backgrounds $\mathbb{RP}^4$ appears to be the minimal  setup. A $\frac{1}{2}$-BPS boundary preserves the same amount of symmetry, but admits many different choices of boundary conditions. By contrast, the case of real projective space is more rigid as conformal inversion does not induce a fixed boundary. In fact,  for $\mathcal{N}=4$ SYM we only need to distinguish two possibilities, namely whether we include charge conjugation $\tau:g\to g^*$, $g\in SU(N)$ in identifying local operators under conformal inversion \cite{Caetano:2022mus}. 

Although straightforward to define as a field theory, constructing the holographic dual of $\mathcal{N}=4$ SYM on $\mathbb{RP}^4$ appears to be rather difficult \footnote{Holography on non-orientable 2d surface has been considered in \cite{Maloney:2016gsg}, and has also been studied in the context of bulk reconstructon in \cite{Miyaji:2015fia,Verlinde:2015qfa,Nakayama:2015mva,Nakayama:2016xvw,Goto:2016wme,Lewkowycz:2016ukf}}. This is because the standard AdS$_5\times$S$^5$ arises as the near-horizon limit of D3 branes in flat space, while the identification for $\mathbb{R}^4\to\mathbb{RP}^4$ is a conformal isometry which only emerges in the low energy limit. Therefore, in lack of a complete string theory picture, the construction of the holographic dual as was done in \cite{Caetano:2022mus} in the supergravity limit, is necessarily bottom-up and involves certain unfixed ingredients. But this suggests that bootstrap strategies, which only rely on symmetries and consistency conditions, may be useful in studying the dual limit. Importantly, \cite{Caetano:2022mus} pointed out that the choice of charge conjugation makes the holographic duals physically very different. Without charge conjugation, the dual is given by a new classical background that is asymptotic AdS$_5\times$S$^5$. With charge conjugation, however, the background is given by just a $\mathbb{Z}_2$ quotient of AdS$_5\times$S$^5$ with an O1 orientifold sitting at the fixed locus S$^2\subset$S$^5$. We will focus on the latter more regular case and compute two-point functions of $\frac{1}{2}$-BPS operators at AdS tree level, i.e., $\mathcal{O}(1/N)$. This corresponds to results in the strongly coupled regime with infinite 't Hooft coupling. We will use a bootstrap algorithm to show that analytic progress is possible even without an explicit effective Lagrangian and determine all two-point functions with arbitrary weights.

\vspace{0.5cm}
\noindent {\bf Kinematics.} We consider $\frac{1}{2}$-BPS operators of the form
\begin{equation}
\mathcal{O}_p(x,Y)=\mathcal{N}_p\,{\rm tr} (\Phi^{i_1}(x)\ldots \Phi^{i_p}(x))Y_{i_1}\ldots Y_{i_p}\;,
\end{equation}
where $\Phi^{i=1,\ldots,6}$ are the six scalars of SYM and $Y_i$ is a null R-symmetry polarization vector with $Y\cdot Y=0$ to ensure the operator is in the rank-$p$ symmetric traceless representation of $SO(6)_R$. The normalization $\mathcal{N}_p$ is such that the two-point function has unit coefficient. Placing the theory on $\mathbb{RP}^4$ breaks the $\mathcal{N}=4$ superconformal symmetry to $OSp(4|4)$  \cite{Wang:2020jgh,Caetano:2022mus}. In particular, only $SO(4,1)\subset SO(4,2)$ survives as the residual conformal symmetry. R-symmetry is also broken into $SO(3)\times SO(3)$ and we can split $Y$ into $Y=(\vec{u},\vec{v})$, with three dimensional vectors $\vec{u}$, $\vec{v}$ satisfying $\vec{u}^2=-\vec{v}^2$. It is also convenient to define $\bar{Y}=(\vec{u},-\vec{v})$. Because of reduced symmetry, one-point functions of scalar operators can be non-vanishing
\begin{equation}\label{Op1pt}
\llangle \mathcal{O}_p\rrangle=a_p\frac{(Y\cdot\bar{Y})^{\frac{p}{2}}}{(1+x^2)^p}\;,\quad p\;\;{\rm even}\;,
\end{equation}
where $\llangle \rrangle$ denotes the nontrivial background $\mathbb{RP}^4$ and $a_p$ is a new CFT data. The unbroken bosonic symmetries also determine the two-point function up to a function of three cross ratios
\begin{equation}
\llangle \mathcal{O}_{p_1}\mathcal{O}_{p_2}\rrangle=\frac{(Y_1\cdot \bar{Y}_1)^{\frac{p_1}{2}}(Y_2\cdot \bar{Y}_2)^{\frac{p_2}{2}}}{(1+x_1^2)^{p_1}(1+x_2^2)^{p_2}}\mathcal{G}_{p_1p_2}(\eta;\sigma,\bar{\sigma})\;,
\end{equation}
where the cross ratios are defined as
\begin{equation}
\eta=\frac{x_{12}^2}{(1+x_1^2)(1+x_2^2)}\;,
\end{equation}
\begin{equation}
\sigma=\frac{Y_1\cdot Y_2}{(Y_1\cdot \bar{Y}_1)^{\frac{1}{2}}(Y_2\cdot \bar{Y}_2)^{\frac{1}{2}}}\;,\; \bar{\sigma}=\frac{Y_1\cdot \bar{Y}_2}{(Y_1\cdot \bar{Y}_1)^{\frac{1}{2}}(Y_2\cdot \bar{Y}_2)^{\frac{1}{2}}}\;.
\end{equation}
Note that the correlator is only nonzero when $p_1+p_2$ is even and   $\mathcal{G}_{p_1p_2}(\eta;\sigma,\bar{\sigma})$ is a polynomial of $\sigma$ and $\bar{\sigma}$ of degree $p_m=\min\{p_1,p_2\}$. Without loss of generality, we will assume $p_m=p_2$ in this paper. The $\mathbb{Z}_2$ quotient also acts on the R-symmetry $Y\to \bar{Y}$. Therefore, the $\mathbb{Z}_2$ invariance of the correlator implies the crossing relation
\begin{equation}
\mathcal{G}_{p_1p_2}(\eta;\sigma,\bar{\sigma})=\mathcal{G}_{p_1p_2}(1-\eta;\bar{\sigma},\sigma)\;.
\end{equation}
Fermionic generators impose extra constraints which are the superconformal Ward identities. Although having not been derived (e.g., from superspace analysis), they can be obtained from the boundary CFT case by analytic continuation. This is based on the observation that $\frac{1}{2}$-BPS boundary condition preserves the same superconformal group $OSp(4^*|4)$ (up to a Wick rotation) and the conformal blocks are essentially identical (up to a minus sign in the cross ratio) \cite{Giombi:2020xah}. Using the BCFT superconformal Ward identities \cite{Liendo:2016ymz}, we find the following condition
\begin{equation}\label{scfWard}
\left(\partial_{w_1}+\frac{1}{2}\partial_z\right)\mathcal{G}_{p_1p_2}(z;w_1,w_2)\bigg|_{w_1=z}=0\;,
\end{equation}
and similarly with $w_1\leftrightarrow w_2$, where we made the change of variables
\begin{equation}
\begin{split}
{}&\eta=-\frac{(z-1)^2}{4z}\;,\\
{}&\sigma=\frac{(1-w_1)(1-w_2)}{4\sqrt{w_1w_2}}\;,\quad \bar{\sigma}=\frac{(1+w_1)(1+w_2)}{4\sqrt{w_1w_2}}\;.
\end{split}
\end{equation}
Note that when we set $w_1=w_2=z$, (\ref{scfWard}) implies the correlator is topological and only depends on $\vartheta={\rm sgn}(z)$
\begin{equation}
\mathcal{G}_{p_1p_2}(z;z,z)=T_{p_1p_2}(\vartheta)\;.
\end{equation}
This follows from the fact that the preserved algebra is isomorphic to 3d $\mathcal{N}=4$ and one can apply the topological twisting of \cite{Chester:2014mea}. 

\begin{figure}
    \centering
   \includegraphics[width=0.4\textwidth]{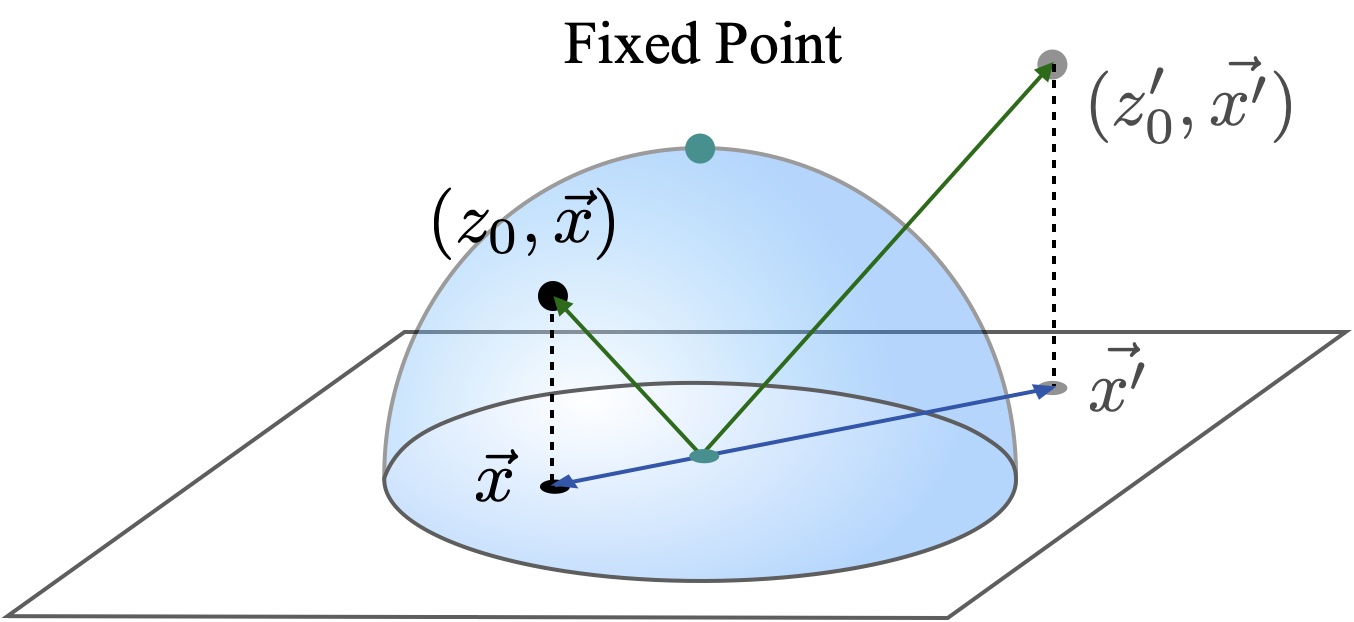}
    \caption{The $\mathbb{Z}_2$ quotient of AdS space in Poincar\'e coordinates where points inside and outside of the unit hemisphere are identified under conformal inversion.}
    \label{fig:quotientAdS}
\end{figure}

\vspace{0.5cm}
\noindent {\bf Quotient AdS and Witten diagrams.} $\mathcal{N}=4$ SYM with gauged charge conjugation is dual to IIB supergravity on AdS$_5\times$S$^5/\mathbb{Z}_2$  \cite{Caetano:2022mus}. In particular, in Poincar\'e coordinates $\mathbb{Z}_2$ acts on AdS$_5$ as
\begin{equation}
\mathcal{I}_{\rm AdS}:\quad(z_0,\vec{z})\to \left(\frac{z_0}{z_0^2+\vec{z}^2},-\frac{\vec{z}}{z_0^2+\vec{z}^2}\right)\;.
\end{equation}
This is a conformal inversion with respect to a unit hemisphere at the origin and leaves invariant the north pole $(1,\vec{0})$.  In fact,  this fixed point extends to a fixed locus S$^2\subset$S$^5$ in the internal space which is the world volume of an O1 orientifold \cite{Caetano:2022mus}. The orientifold is needed because the five-form flux vanishes unless the worldsheet orientation is also reversed. The orientifold effective action on S$^2$ induces vertices in AdS$_5$ which have support only at a point.

\begin{figure}
    \centering
   \includegraphics[width=0.29\textwidth]{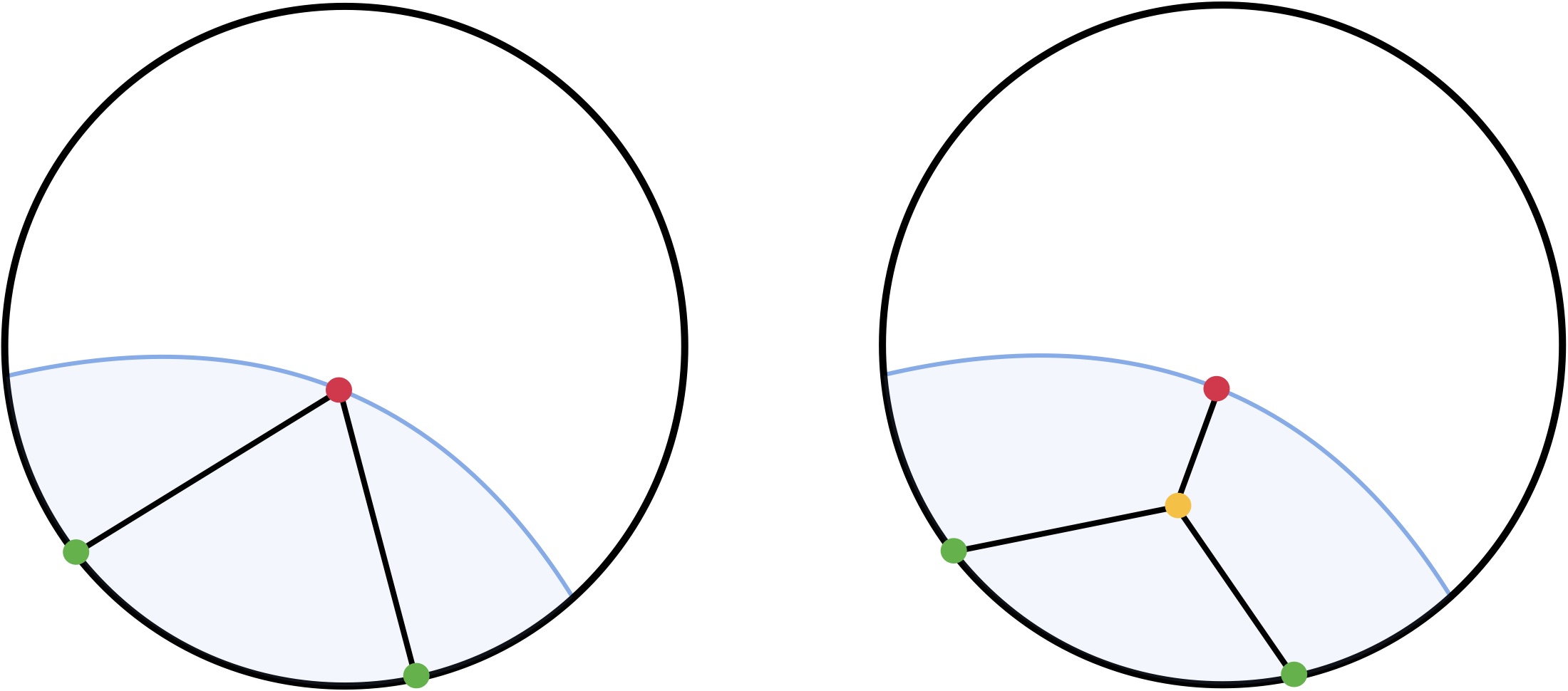}
    \caption{Contact and exchange Witten diagrams.}
    \label{fig:Wittendiagrams}
\end{figure}

\begin{figure}
    \centering
   \includegraphics[width=0.45\textwidth]{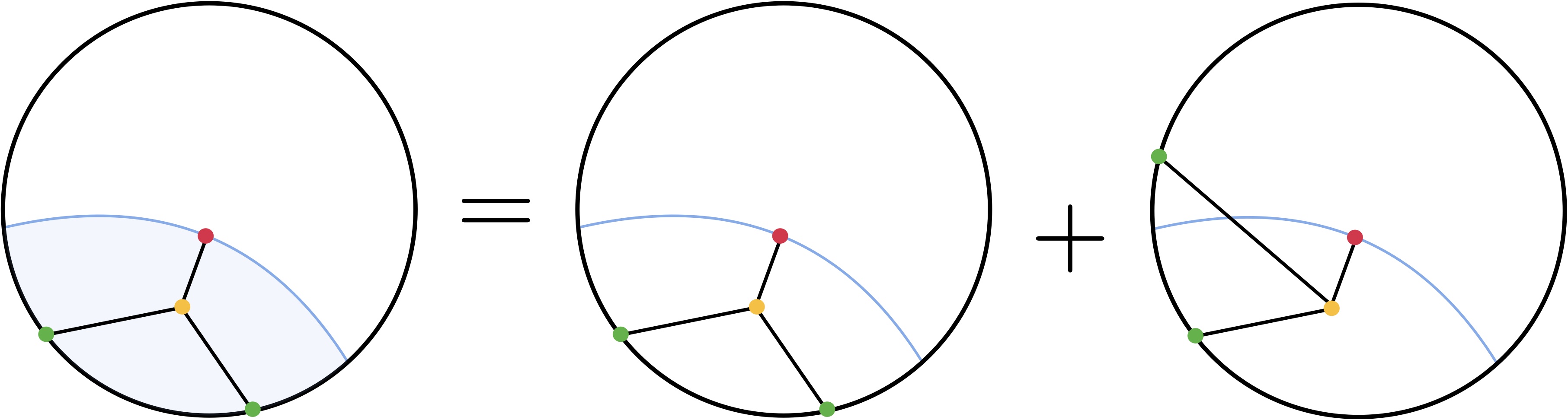}
    \caption{Method of images for exchange Witten diagrams.}
    \label{fig:exchWDintoWWb}
\end{figure}

This leads to two types of Witten diagrams at tree level (Fig. \ref{fig:Wittendiagrams}). The simpler contact Witten diagram is just a product of bulk-to-boundary propagators with the bulk point anchored at the fixed point. Writing it in terms of the cross ratio, we have
\begin{equation}
\mathcal{W}_{\rm con}=1\;.
\end{equation}
Because symmetry dictates that only scalar can have non-vanishing one-point function, all internal lines in exchange Witten diagrams must be scalar fields. For IIB supergravity, this narrows the fields down to those in Table \ref{tab:componentfields} where $\mathcal{O}_p$ is dual to $s_p$. A scalar exchange Witten diagram in AdS$_{d+1}/\mathbb{Z}_2$ can be obtained by the method of images and is the sum of the exchange Witten diagram in full AdS and its inversion image \cite{Giombi:2020xah} (see Fig. \ref{fig:exchWDintoWWb})\begin{equation}
\mathcal{W}^{\mathbb{Z}_2}_\Delta(\eta)=\mathcal{W}_\Delta(\eta)+\bar{\mathcal{W}}_\Delta(\eta)\;.
\end{equation}
Here the image Witten diagram is related by $\bar{\mathcal{W}}_\Delta(\eta)=\mathcal{W}_\Delta(1-\eta)$. When the internal dimension $\Delta=p$ satisfies $p_1+p_2-p\in 2\mathbb{Z}_+$, the exchange Witten diagram truncates to a rational function of $\eta$ \cite{Giombi:2020xah}
\begin{equation}\label{Wp}
\begin{split}
\mathcal{W}_p={}&\sum_{k=\frac{p-p_1-p_2}{2}}^{-1} \frac{\Gamma(1-\frac{d}{2}+p)\Gamma(k+p_1)\Gamma(k+p_2)}{\Gamma(\frac{p+p_{12}}{2})\Gamma(\frac{p-p_{12}}{2})}\\
{}&\quad \times  \frac{\eta^{k}}{\Gamma(\frac{2+2k-p+p_1+p_2}{2})\Gamma(\frac{2-d+2k+p+p_1+p_2}{2})}\;,
\end{split}
\end{equation}
where $p_{12}=p_1-p_2$. Here we normalized the exchange Witten diagrams such that under conformal block decomposition the single-trace conformal block has unit coefficient. For the ``extremal'' cases with $p_1+p_2=p$ or $p_2+p=p_1$, we define $\mathcal{W}_p$ to be zero. This is because such cubic vertices would lead to divergent three-point functions in the unquotiented theory and must be absent \cite{Rastelli:2016nze,Rastelli:2017udc}. 

\begin{table}
\begin{center}
\begin{tabular}{|c|c|c|c|}\hline fields & $s_p$ & $\phi_p$ & $t_p$ \\\hline $SU(4)$ irrep & $[0,p,0]$ & $[2,p-4,2]$ & $[0,p-4,0]$ \\\hline $\Delta$ & $p$ & $p+2$ & $p+4$ \\\hline \end{tabular}
\caption{Allowed fields in exchange Witten diagrams.}\label{tab:componentfields}
\end{center}
\end{table}

\vspace{0.5cm}
\noindent {\bf Bootstrap.} We will compute tree-level two-point functions using a bootstrap strategy. This allow us to see to which extent these observables are fixed by symmetries. We start with the following ansatz, in a similar fashion as in \cite{Rastelli:2016nze,Rastelli:2017udc,Gimenez-Grau:2023fcy,Chen:2023yvw}
\begin{equation}
\mathcal{A}_{p_1p_2} =\mathcal{A}_{p_1p_2,e} +\bar{\mathcal{A}}_{p_1p_2,e}+\mathcal{A}_{p_1p_2,c}\;,
\end{equation}
where $\mathcal{A}_{p_1p_2,e}$ includes all the exchange Witten diagrams in the direct channel with unfixed parameters $\lambda_X$
\begin{equation}\label{ansatzAe}
\mathcal{A}_{p_1p_2,e}(\eta,\sigma,\bar{\sigma}) =\sum_X \lambda_X h_{R_X}(\sigma,\bar{\sigma}) \mathcal{W}_{\Delta_X}(\eta)\;,
\end{equation}
and $\bar{\mathcal{A}}_{p_1p_2,e}$ is its inversion mirror
\begin{equation}
\bar{\mathcal{A}}_{p_1p_2,e}(\eta,\sigma,\bar{\sigma})= \mathcal{A}_{p_1p_2,e}(1-\eta,\bar{\sigma},\sigma)\;.
\end{equation}
The allowed fields $X$ are selected from Table \ref{tab:componentfields} based on two criteria
\begin{enumerate}
 \item R-symmetry selection rule: the R-symmetry representation of $X$ must appear in  $[0,p_1,0]\times [0,p_2,0]$. 
 \item Non-extremal: $\Delta_X<p_1+p_2$ and $p_2+\Delta_X<p_1$. As mentioned, this is to avoid divergent Witten diagrams.
\end{enumerate} 
It is easy to see that the spectrum always satisfies the truncation condition so all exchange Witten diagrams can be evaluated using (\ref{Wp}).
The R-symmetry information of the exchanged fields is encoded in the polynomials $h_{R_X}(\sigma,\bar{\sigma})$  which are solutions to the R-symmetry Casimir equation. Only two types of  representations with $SU(4)$ Dynkin labels $[0,p,0]$ and $[2,p,2]$ are relevant and the corresponding polynomials $h_{p,0}$ and $h_{p,2}$ are given explicitly in (\ref{Rsymmpoly}) of the Appendix. For the contact part $\mathcal{A}_{p_1p_2,c}$, we assume that the contact vertices have no derivatives but all possible R-symmetry structures
\begin{equation}\label{ansatzcon}
\mathcal{A}_{p_1p_2,c}=\sum_{a=0}^{p_m}\sum_{b=0}^{p_m-a}\delta_{a+b-p_m,{\rm even}}c_{ab}\sigma^a\bar{\sigma}^b\;. 
\end{equation}
Finally, we impose the superconformal Ward identities (\ref{scfWard}) to solve for the coefficients. 

Let us implement this strategy starting with small values of $p_m$. For $p_m=p_2=2$, only $s_{p_1}$ can be exchanged. We find that (\ref{scfWard}) fixes all contact coefficients in terms of the coefficient of $s_{p_1}$, except for an additive constant which solves (\ref{scfWard}) trivially. This gives
\begin{equation}
\mathcal{G}_{p_12}=\mu_1\frac{\sigma\bar{\sigma}}{\eta(1-\eta)}+\mu_0\;,
\end{equation}
with unfixed parameters $\mu_0$ and $\mu_1$. We then consider $p_m=p_2=3$. In this case, $\{X\}=\{s_{p_1-1},s_{p_1+1}\}$. Solving the Ward identities completely fixes the contact terms where we note there is no constant term in (\ref{ansatzcon}) for odd $p_m$. It also determines the relative coefficient of $s_{p_1-1}$ and $s_{p_1+1}$
\begin{equation}
\frac{\lambda_{s_{p_1-1}}}{\lambda_{s_{p_1+1}}}=\frac{p_1-1}{p_1+1}\;.
\end{equation}
Note in our normalizations $\lambda_{s_p}$ is the product of three-point and one-point function coefficients $\lambda_{s_p}=C_{p_1p_2p}a_p$, with $C_{p_1p_2p}$ obtained in \cite{Lee:1998bxa} as $C_{p_1p_2p}=\sqrt{p_1p_2p}$.
Therefore, we can solve the relation and get
\begin{equation}\label{1ptap}
a_p=C\sqrt{p}\;,
\end{equation}
where $C$ is an overall constant independent of $p$ \footnote{This result for the one-point function coefficients also appears to be consistent with the integrability result \cite{kw}.}. We can similarly proceed to higher values of $p_m$ where $\phi_p$ and $t_\phi$ also start appearing. If we start with (\ref{ansatzAe}), we find an increasing number of unfixed coefficients. Quite remarkably, further imposing the condition (\ref{1ptap}), as is required by $p_m=3$, solves all coefficients in $\llangle \mathcal{O}_{p_1} \mathcal{O}_{p_2} \rrangle$ up to an additive constant $B_{p_1p_2}$ when $p_m$ is even. The general solution is
\begin{equation}\label{gen2pt}
\begin{split}
\mathcal{G}_{p_1p_2}={}&\bigg\{\sum_{p\in\mathcal{S}} a_pC_{p_1p_2p}\big(h_{p,0}\mathcal{W}_p+\nu_ph_{p-4,2}\mathcal{W}_{p+2}\\
+{}&\rho_p h_{p-4,0}\mathcal{W}_{p+4} \big)+(\eta\to1-\eta\,,\,\sigma\leftrightarrow\bar{\sigma})\bigg\}\\
+{}&\mathcal{G}_{p_1p_2,{\rm con}}+B_{p_1p_2} \delta_{p_m,{\rm even}}\;,
\end{split}
\end{equation}
where 
\begin{equation}\label{Gp1p2contact}
\begin{split}
{}&\mathcal{G}_{p_1p_2,{\rm con}}=C\sum_{t=1}^{\lfloor\frac{p_m}{2}\rfloor}\bigg[\frac{(\frac{1+p_m}{2}-t)_t(\frac{2+p_m}{2}-t)_t\Gamma(\frac{3-p_1-p_2}{2})}{\Gamma(t)\Gamma(\frac{p_{12}}{2}+t)}\\
{}&\quad\quad\times\frac{4(-1)^{\lfloor\frac{p_m-1}{2}\rfloor}\sqrt{p_1}(\lfloor\frac{1+p_m}{2}\rfloor)_{\frac{p_{12}}{2}}}{\sqrt{p_m}(\lfloor\frac{1+p_m}{2}\rfloor)_{\frac{1-p_1-p_2}{2}}}(\sigma+\bar{\sigma})^{p_m-2t}\bigg]\;,
\end{split}
\end{equation}
and the coefficients are given by 
\begin{equation}
\begin{split}
\nu_p={}&\frac{((p-2)^2-p_{12}^2)(p^2-p_{12}^2)}{16(p-3)(p-1)^2p}\;,\\
\rho_p={}&\frac{(p-1)((p+2)^2-p_{12}^2)(p^2-p_{12}^2)}{16(p^2-4)p(p+1)^2}\nu_p\;. 
\end{split}
\end{equation}
The sum over $p$ is restricted by the selection rules to be in the set $\mathcal{S}=\{|p_{12}|+2,|p_{12}|+4,\ldots, p_1+p_2-2\}$, where we also recall that $\mathcal{W}_p$ is defined to be zero if $p\geq p_1+p_2$. The constants $B_{p_1p_2}$ are not fixed by the bootstrap alone. But as we will see, the supergravity analysis  strongly indicates that they are absent.

\vspace{0.5cm}
\noindent {\bf Supergravity calculation.} The effective action of the O1 orientifold includes a tension term
\begin{equation}
-T_{\rm O1}\int_{{\rm S}^2}e^{-\frac{\Phi}{2}}\sqrt{-\det g_{ab}^{\rm P.B.}}\;,
\end{equation}
with $T_{\rm O1}\sim N/\sqrt{\lambda}$, and a Wess-Zumino term which is not needed here. Expanding the fields around the background and integrating over S$^2$ gives rise to vertices localized at the fixed point. For tree-level two-point functions of $\mathcal{O}_p$, the relevant fluctuations are those of the metric $g\to \bar{g}+h$. In particular, it involves only the combination $\pi=h^\alpha{}_\alpha$ with $\alpha$ along S$^5$, which can be decomposed into spherical harmonics and is related to $s$ and $t$ by \cite{Kim:1985ez,Arutyunov:1998hf} \footnote{Here we have changed the labelling of $t$ in contrast to the literature to agree with Table \ref{tab:componentfields}.}
\begin{eqnarray}
\pi(x,y)&=&\sum \pi_p^I(x)Y_p^I(y)\;,\\
\pi_p&=&10ps_p+10(p+4)t_{p+4}\label{pist}\;.
\end{eqnarray}
The expansion leads to terms linear and quadratic in $\pi$, as well as terms of higher order which are irrelevant at tree level. From the linear term $\pi$, we can easily reproduce $a_p\propto \sqrt{p}$ using standard Witten diagram prescriptions and properties of spherical harmonics (see Appendix for details). We can similarly compute the one-point function of the operator dual to $t_{p+4}$ and find $a_{t_{p+4}}\propto \sqrt{\frac{(p+3)(p+4)(p+7)}{(p+1)(p+5)}}$, which gives the correct ratio $\rho_p$ after  including the three-point integral \cite{Arutyunov:1999en}. On the other hand, contact terms are more subtle. They can come from the quadratic term $\pi^2$. Its   contribution is 
\begin{equation}\label{Pip1p2}
\begin{split}
{}&\Pi_{p_1p_2}=\frac{25\sqrt{p_1p_2}(p_1+1)(p_2+1)}{2}\sum_{p\in\mathcal{I}} \frac{4\pi}{(p_1+p_2+1)!!}\\
{}&\quad\times \binom{p_1}{p}\binom{p_2}{p} p! (p_1-p-1)!!(p_2-p-1)!! (\sigma+\bar{\sigma})^p\;, 
\end{split}
\end{equation}
where $\mathcal{I}=\{p_m,p_m-2,\ldots,\frac{1-(-1)^{p_m}}{2}\}$ and the expression has a combinatoric meaning (see Appendix). Although this does not reproduce (\ref{Gp1p2contact}), it is related in a simple way
\begin{equation}
\begin{split}
\mathcal{G}_{p_1p_2,{\rm con}}={}&C\frac{(p_1+p_2+1)(p_1+p_2-1)}{50\pi(p_1+1)(p_2+1)p_1p_2}\\
{}&\times(\Sigma\partial_\Sigma-p_1)(\Sigma\partial_\Sigma-p_2)\Pi_{p_1p_2}\;,
\end{split}
\end{equation} 
where $\Sigma=\sigma+\bar{\sigma}$. Note that the disagreement does not disprove the bootstrap result because (\ref{Pip1p2}) violates superconformal symmetry. Instead, it suggests that the supergravity analysis is more subtle and there are other unknown sources of contact terms \footnote{For example, they can come from boundary terms which are ignored when we use the equation of motion.}. A similar failure of the naive supergravity calculation was also pointed out in \cite{Gimenez-Grau:2023fcy} for the case of Wilson loops. Nevertheless, the calculation captures two essential features of contact contributions arising from integrating over S$^2$. The first is that R-symmetry cross ratios enter only via the combination $\sigma+\bar{\sigma}$. The second is analyticity in its power $p$. Note the ambiguities $B_{p_1p_2}$ correspond to $p=0$. Although we have not exactly reproduced the contact terms from supergravity, analyticity strongly suggests that we should set $B_{p_1p_2}$  to zero in (\ref{gen2pt}).

\vspace{0.5cm}
\noindent {\bf Discussions.} In this paper we studied connected two-point functions in $\mathcal{N}=4$ SYM on $\mathbb{RP}^4$ in the strongly coupled regime. The bootstrap techniques allowed us to determine them at leading order in $1/N$ for $\frac{1}{2}$-BPS operators with arbitrary weights up to a common overall coefficient and an additive constant for each correlator. The results are very simple and are just rational functions of cross ratios. Using analyticity we further argued that these additive constants should be set to zero. However, it would be desirable to independently check this using other methods. One option is supersymmetric localization, which in the context of  $\mathcal{N}=4$ SYM on $\mathbb{RP}^4$ has been initiated in \cite{Wang:2020jgh}. Along the line of \cite{Pufu:2023vwo,Billo:2023ncz,Dempsey:2024vkf,Billo:2024kri}, where superconformal line defects were studied, we can compute integrated two-point functions using localization. This will be sufficient to fix the overall constant and to determine the additive ambiguities. It would also be very interesting to perform cross checks using integrability methods where the integrability of this setup was pointed out in \cite{Caetano:2022mus} and will be discussed in \cite{kw}. 

There are several other interesting directions to explore. The first is trying to identify possible hidden structure of higher dimensional conformal symmetry \cite{Caron-Huot:2018kta,Rastelli:2019gtj,Alday:2021odx,Zhou:2021gnu,Abl:2021mxo}. This hidden symmetry appears to be associated with the conformal flatness of the background, which is also the case of AdS$_5\times$S$^5/\mathbb{Z}_2$. Finding its explicit realization will help us understand better the nature of hidden conformal symmetry, as well as find similar structures in the Wilson loop case \cite{Gimenez-Grau:2023fcy}. The second direction is to go to higher orders in $1/N$ which correspond to loop corrections in AdS. At one-loop order, this should be possible to achieve by using the AdS unitarity method \cite{Aharony:2016dwx} which has been recently generalized to the closely related defect case in \cite{Chen:2024orp}. Another exciting future avenue is to study stringy corrections (i.e., $1/\lambda$ corrections) to two-point functions by using a combination of localization and integrability techniques. This provides an attractive alternative for similar investigations in four-point functions (see, e.g., \cite{Binder:2019jwn,Chester:2019jas,Chester:2020vyz,Alday:2022uxp,Alday:2023jdk,Alday:2023mvu,Brown:2023zbr}) and defect two-point functions \cite{Pufu:2023vwo,Billo:2023ncz,Dempsey:2024vkf,Billo:2024kri} due to its simplicity. Related to this, it would be very interesting to understand how to take the flat-space limit of two-point functions and compare with flat-space calculations, although a suitable formalism similar to Mellin space \cite{Mack:2009mi,Penedones:2010ue,Rastelli:2017ecj,Goncalves:2018fwx} perhaps needs to be established first. Finally, it would also be interesting to compute two-point functions in the other setup without gauging charge conjugation. This case appears more challenging because of the highly nontrivial geometry \cite{Caetano:2022mus}. But one can presumably make progress for the lowest KK mode in a way similar to \cite{Chiodaroli:2016jod}. 

\vspace{0.5cm}
\noindent {\bf Acknowledgements.} We thank Konstantinos Rigatos, Minwoo Suh, Yifan Wang for very helpful discussions, Konstantinos Rigatos and Minwoo Suh for collaboration during an early stage of this project, and Aleix Gimenez-Grau and Yifan Wang for collaborating on a related project. We also thank Joao Caetano, Shota Komatsu and Leonardo Rastelli for correspondence, and are  grateful to Shota Komatsu, Konstantinos Rigatos and Yifan Wang for useful comments on draft. This work is supported by the NSFC Grant No. 12275273, funds from Chinese Academy of Sciences, University of Chinese Academy of Sciences, and the Kavli Institute for Theoretical Sciences. The work of X.Z. is also supported by the NSFC Grant No. 12247103 and the Xiaomi Foundation. X.Z. thanks the warm hospitality of his alma mater university during his visit where part of the work was done.

\appendix

\section{R-symmetry polynomials}
The R-symmetry polynomials satisfy the following Casimir equation
\begin{equation}
\begin{split}
\nonumber{}&\bigg[\sigma^2(\bar{\sigma}^2-\sigma^2+1)\partial^2+(\bar{\sigma}^2-1)^2\bar{\partial}^2+\bar{\sigma}(\bar{\sigma}^2-\sigma^2-1)\bar{\partial}\\
{}&-\sigma(\sigma^2-\bar{\sigma}^2+3)\partial-\sigma^2(\bar{\sigma}^2+1)\bar{\partial}^2-2\sigma\bar{\sigma}(\sigma^2-\bar{\sigma}^2+1)\partial\bar{\partial}\\
{}&+\frac{p_{12}^2}{4}(\sigma^2-\bar{\sigma}^2+1)-\frac{1}{2}C_{p,q}\bigg]\frac{h_{p,q}(\sigma,\bar{\sigma})}{\sigma^{\frac{p_1+p_2}{2}}}=0\;,
\end{split}
\end{equation}
where $C_{p,q}=(q+2)(p+2q)+(p+2)(p+q)$ is the Casimir eigenvalue. Let us make the change of variables
\begin{equation}
\frac{w_1+1}{w_1-1}=\mathrm{z}^{-\frac{1}{2}}\;,\quad \frac{w_2+1}{w_2-1}=\bar{\mathrm{z}}^{-\frac{1}{2}}\;,
\end{equation}
and define 
\begin{equation}
\tilde{h}_{p,q}(\mathrm{z},\bar{\mathrm{z}})=\sigma^{-\frac{p_1+p_2-1}{2}}\bar{\sigma}^{\frac{p_1-p_2-1}{2}} h_{p,q}(\sigma,\bar{\sigma})\;.
\end{equation}
It is not difficult to show that the R-symmetry Casimir equation for $\tilde{h}_{p,q}(\mathrm{z},\bar{\mathrm{z}})$ becomes the Casimir equation for four-point conformal blocks in 3d upon identifying
\begin{equation}
\nonumber\Delta_{12}=\frac{1-p_{12}}{2}\;,\; \Delta_{34}=\frac{1+p_{12}}{2}\;,\; \Delta=\frac{1-p-q}{2}\;,\; \ell=\frac{q}{2}\;,
\end{equation}
where $\Delta_{ij}=\Delta_i-\Delta_j$. The functions $\tilde{h}_{p,q}(\mathrm{z},\bar{\mathrm{z}})$ are then the conformal block where $(\mathrm{z},\bar{\mathrm{z}})$ are identified with the standard conformal cross ratios $(z,\bar{z})$. These conformal blocks have been obtained in \cite{Dolan:2003hv} and are most convenient to write down using Jack polynomials 
\begin{equation}
P_{\lambda_1,\lambda_2}(\mathrm{z},\bar{\mathrm{z}})=(\mathrm{z}\bar{\mathrm{z}})^{\frac{\lambda_1+\lambda_2}{2}}C_{\lambda_1-\lambda_2}^{\frac{1}{2}}\left(\frac{\mathrm{z}+\bar{\mathrm{z}}}{2\sqrt{\mathrm{z}\bar{\mathrm{z}}}}\right)\;,
\end{equation}
Here $C_n^k(x)$ is the Gegenbauer polynomial. We are interested in the following two special cases
\begin{equation}\label{Rsymmpoly}
\begin{split}
{}&\tilde{h}_{2n+p_{12},0}=\sum_{0\leq b\leq a\leq \lfloor\frac{n}{2}\rfloor} r_{ab} P_{\frac{1-p_{12}-2n}{4}+a,\frac{1-p_{12}-2n}{4}+b}\;,\\
{}&\tilde{h}_{2n+p_{12},2}=\sum_{0\leq b\leq a+1\leq \lfloor\frac{n+2}{2}\rfloor} s_{ab} P_{\frac{1-p_{12}-2n}{4}+a,\frac{-3-p_{12}-2n}{4}+b}\;,
\end{split}
\end{equation}
where 
\begin{equation}
r_{ab}=\frac{(2 a-2 b+1)\left(\frac{1-n}{2}\right)_a \left(-\frac{n}{2}\right)_a \left(-\frac{n+1}{2}\right)_b \left(-\frac{n}{2}\right)_b}{2 b! \left(\frac{1}{2}\right)_{a+1}\left(\frac{-2 n-p_{12}+1}{2}\right)_a \left(\frac{-2 n-p_{12}}{2}\right)_b}\;,
\end{equation}
\begin{equation}
\begin{split}
s_{ab}={}&\frac{(2 a-2 b+3) (2 n+p_{12}+4) }{2 (2 n+p_{12}+5) (-2 b+2 n+p_{12}+4)b!}\\
{}&\times ((a+b+1) (2 n+p_{12})-4 a b+5 a-b+5)\\{}&\times \frac{\left(\frac{1-n}{2}\right)_a \left(-\frac{n}{2}\right)_a \left(-\frac{n+3}{2}\right)_b \left(-\frac{n+2}{2}\right)_b}{\left(\frac{1}{2}\right)_{a+2} \left(\frac{-2 n-p_{12}+1}{2}\right)_a \left(-\frac{2n+p_{12}+4}{2}\right)_b}\;.
\end{split}
\end{equation}
Note that the special choice of quantum numbers makes the sums over $a$ and $b$ truncating, as compared to the generic case where the sums are infinite. We have also normalized $\tilde{h}_{2n+p_{12},0}$ such that the exchanged representation appears with unit coefficient.

\section{Details of the supergravity calculation}
For simplicity, we focus on the fields $s$ and $t$ which involve only scalar spherical harmonics
\begin{equation}
\nonumber s(x,y)=\sum s^I_p(x) Y_p^I(y)\;,\;\;t(x,y)=\sum t^I_{p+4}(x) Y_p^I(y)\;.
\end{equation}
Here the spherical harmonics are defined by
\begin{equation}
Y^I=\frac{1}{\sqrt{z(p)}}C^I_{i_1\ldots i_p}y^{i_1}\ldots y^{i_p}\;,
\end{equation}
with S$^5$ coordinates $y^{1,\ldots,6}$ satisfying $y\cdot y=1$ and
\begin{equation}
z(p)=\frac{\pi^3}{2^{p-1}(p+1)(p+2)}\;.
\end{equation}
We have also defined an invariant tensor $C^I_{i_1\ldots i_p}$ which satisfies
\begin{eqnarray}
&&C^I_{i_1\ldots i_p}C^J_{i_1\ldots i_p}=\delta^{IJ}\;,\label{CICJ}\\
&&C^I_{i_1\ldots i_p}C^I_{j_1\ldots j_p}=\delta^{i_1\ldots i_p,(j_1\ldots j_p)}+(\text{mixing $i$, $j$})\;.\label{CICI}
\end{eqnarray}
With this normalization, we have
\begin{equation}
\int_{{\rm S}^5}Y^{I_1}_p Y^{I_2}_p=\delta^{I_1I_2}\;.
\end{equation}
The quadratic actions of $s$ and $t$ are \cite{Arutyunov:1998hf}
\begin{equation}\nonumber
\begin{split}
S(s)={}&\int d^5x \sqrt{-g}\xi_{s,p}\left(-\frac{1}{2}\nabla_a s_p\nabla^a s_p-\frac{1}{2}p(p-4)s_p^2\right)\;,\\
S(t)={}&\int d^5x \sqrt{-g}\xi_{t,p}\left(-\frac{1}{2}\nabla_a t_p\nabla^a t_p-\frac{1}{2}p(p+4)t_p^2\right)\;,
\end{split}
\end{equation}
where 
\begin{equation}\nonumber
\xi_{s,p}=\frac{128N^2p(p-1)(p+2)}{(2\pi)^5(p+1)}\;,\;\xi_{t,p}=\frac{128N^2(p-2)p(p+1)}{(2\pi)^5(p-1)}\;.
\end{equation}
On the other hand, for a scalar field normalized as 
\begin{equation}
\int \sqrt{g} \frac{1}{2}(\nabla^\mu \phi\nabla_\mu\phi+p(p-d)\phi^2)\;,
\end{equation}
the bulk-to-boundary propagator is 
\begin{equation}
G_{B\partial}(x,z)=N^{\rm prop}_p\left(\frac{z_0}{z_0^2+(\vec{z}-\vec{x})^2}\right)^p\;,
\end{equation}
where 
\begin{equation}
N^{\rm prop}_p=\frac{\Gamma(p)}{\pi^{\frac{d}{2}}\Gamma(p-d/2)}\;.
\end{equation}
It gives rise to the two-point function \cite{Freedman:1998tz}
\begin{equation}
\langle \mathcal{O}_p(x) \mathcal{O}_p(y)\rangle=\frac{N^{\rm 2pt}_p}{(x-y)^{2p}}\;,
\end{equation}
where 
\begin{equation}
N^{\rm 2pt}_p=\frac{\Gamma(p)(2p-d)}{\pi^{\frac{d}{2}}\Gamma(p-d/2)}\;.
\end{equation}

Let us assume that the linear term in the S$^2$ effective Lagrangian is $\kappa\pi$ where $\kappa\sim N/\sqrt{\lambda}$. To compute the one-point function coefficient of $\mathcal{O}_p$, we also need to translate to our index-free notation. This is achieved by using (\ref{CICI})
\begin{equation}
\llangle \mathcal{O}_p(x,Y)\rrangle=\frac{1}{p!}\llangle \mathcal{O}_p^I (x) C^I_{i_1\ldots i_p} \rrangle Y^{i_1}\ldots Y^{i_p}\;.
\end{equation}
Recalling (\ref{pist}), we find that the one-point function is given by the Witten diagram as  
\begin{equation}\nonumber
\llangle \mathcal{O}_p(x,Y)\rrangle=10\kappa p \xi_{s,p}^{-\frac{1}{2}} N^{\rm prop}_p (N^{\rm 2pt}_p)^{-\frac{1}{2}}\frac{1}{\sqrt{z(p)}}\frac{Z_p}{(1+x^2)^p}\;,
\end{equation}
where we still need to perform the integral
\begin{equation}\nonumber
Z_p=\frac{1}{p!}\int_{{\rm S}^2} C^I_{i_1\ldots i_p}y^{i_1}\ldots y^{i_p}C^I_{j_1\ldots j_p}Y^{j_1}\ldots Y^{j_p}\;.
\end{equation}
It is important to note that since the integral is restricted to S$^2$, only the $i=1,2,3$ components of $y^i$ are nonzero. The integral gives pairwise Kronecker deltas among the $i$ indices and then among $j$ via (\ref{CICI}). The restriction on $i$ is then carried over to $j$ and projects $Y$ to its first three components. This allows us to write down invariants $\sum_{j=1}^3 Y^j Y^j=\frac{1}{2}(Y\cdot \bar{Y})$. It is easy to see that the integral is only nonzero when $p$ is even and is given by
\begin{equation}
Z_p=\frac{2^{2-\frac{p}{2}}\pi}{p+1} (Y\cdot \bar{Y})^{\frac{p}{2}}\delta_{p,{\rm even}}\;.
\end{equation}
Comparing with (\ref{Op1pt}) we find
\begin{equation}
a_p=\frac{10\pi\kappa}{N} \sqrt{p}\;.
\end{equation}
The analysis for the one-point function of $t_p$ is similar and gives 
\begin{equation}\nonumber
\begin{split}
a_{t_{p+4}}={}&10\kappa(p+4)\xi_{t,p+4}^{-\frac{1}{2}}N^{\rm prop}_{p+8} (N^{\rm 2pt}_{p+8})^{-\frac{1}{2}}\frac{Z(p)}{\sqrt{z(p)}}\\
={}&\frac{10\pi\kappa}{N}\sqrt{\frac{(p+3)(p+4)(p+7)}{(p+1)(p+5)}}\;.
\end{split}
\end{equation}

Let us now assume a quadratic term $\kappa' \pi^2$ in the effective Lagrangian with $\kappa'$ the same order as $\kappa$. We can similarly compute its contribution to the contact part
\begin{equation}\nonumber
\begin{split}
{}&\llangle \mathcal{O}_{p_1}(x_1,Y_1) \mathcal{O}_{p_2}(x_2,Y_2)\rrangle_{\pi^2}=200\kappa' p_1p_2\xi_{s,{p_1}}^{-\frac{1}{2}}\xi_{s,p_2}^{-\frac{1}{2}}\\
{}&\times \frac{N^{\rm prop}_{p_1}N^{\rm prop}_{p_2} (N^{\rm 2pt}_{p_1}N^{\rm 2pt}_{p_2})^{-\frac{1}{2}}}{(1+x_1^2)^{p_1}(1+x_2^2)^{p_2}}\frac{1}{\sqrt{z(p_1)z(p_2)}}L_{p_1p_2}\;,
\end{split}
\end{equation}
where $L_{p_1p_2}$ is the S$^2$ integral
\begin{equation}\nonumber
\begin{split}
L_{p_1p_2}={}&\frac{1}{p_1!p_2!}\int_{{\rm S}^2}C^I_{i_1\ldots i_{p_1}}C^J_{j_1\ldots j_{p_2}}y^{i_1}\ldots y^{i_{p_1}}y^{j_1}\ldots y^{j_{p_2}}\\
{}&\times C^I_{k_1\ldots k_{p_1}}C^J_{l_1\ldots l_{p_2}}Y_1^{k_1}\ldots Y_1^{k_{p_1}}Y_2^{l_1}\ldots Y_2^{l_{p_2}}\;.
\end{split}
\end{equation}
Again, the integration gives Wick contractions among $y$. A contraction between two $y_i$ or two $y_j$ leads to $\frac{1}{2}Y_1\cdot \bar{Y}_1$ or $\frac{1}{2}Y_2\cdot \bar{Y}_2$ respectively. A contraction between one $y_i$ and one $y_j$ leads to instead $\frac{1}{2}(Y_1\cdot Y_2+Y_1\cdot \bar{Y}_2)$. Therefore, we find
\begin{equation}\nonumber
\begin{split}
{}&L_{p_1p_2}=(Y_1\cdot\bar{Y}_1)^{\frac{p_1}{2}}(Y_2\cdot\bar{Y}_2)^{\frac{p_2}{2}}\frac{4\pi}{(p_1+p_2+1)!!}2^{-\frac{p_1+p_2}{2}} \\
{}&\times \sum_{p\in\mathcal{I}}\binom{p_1}{p}\binom{p_2}{p} p! (p_1-p-1)!!(p_2-p-1)!! (\sigma+\bar{\sigma})^p\;,
\end{split}
\end{equation}
where the coefficient in the summand counts the number of such a partition of $p_1$, $p_2$. Altogether, we get 
\begin{equation}
\llangle \mathcal{O}_{p_1} \mathcal{O}_{p_2}\rrangle_{\pi^2}=\frac{\kappa'}{N^2} \frac{(Y_1\cdot \bar{Y}_1)^{\frac{p_1}{2}}(Y_2\cdot \bar{Y}_2)^{\frac{p_2}{2}}}{(1+x_1^2)^{p_1}(1+x_2^2)^{p_2}}\Pi_{p_1p_2}\;,
\end{equation}
where $\Pi_{p_1p_2}$ was defined in (\ref{Pip1p2}).

\bibliography{refs} 
\bibliographystyle{utphys}
\end{document}